\documentstyle[aps,prb,floats,epsfig]{revtex}
\begin{document}
\title{Non-equilibrium dynamics in an interacting nanoparticle system}
\author{P. J{\"o}nsson$^1$, M. F. Hansen$^{1,2}$, and P. Nordblad$^1$}
\address{
1. Department of Materials Science, Uppsala University\\
        Box 534, SE-751 21 Uppsala, Sweden\\
2. Department of Physics, Building 307, Technical University of
Denmark\\
DK-2800 Lyngby, Denmark}

\date{\today}
\maketitle

\begin{abstract}
Non-equilibrium dynamics in an interacting Fe-C nanoparticle sample,
exhibiting a low temperature spin glass like phase, has been studied
by low frequency ac-susceptibility and magnetic relaxation
experiments.
The non-equilibrium behavior shows characteristic
spin glass features, but some qualitative differences exist.
The nature of these differences is discussed.

\end{abstract}

\section{Introduction}
The dynamics of interacting nanoparticle systems has been subject of
considerable interest concerning the existence of a low temperature
spin glass phase.
Evidence for a phase transition from a high temperature
superparamagnetic phase to a low temperature spin glass like phase
has been given from reports on critical slowing down \cite{Djurberg} and a
divergent behavior of the nonlinear susceptibility.
\cite{Jonsson,Mamiya}

Spin glasses have been intensively investigated during the two last
decades, not only concerning the phase transition but also the nature
of the spin glass phase.  The non-equilibrium dynamics in the spin
glass phase has been extensively studied by conventional and also
more sophisticated dc-magnetic relaxation and low frequency
ac-susceptibility experiments.\cite{SGrev} A non-equilibrium character
of the low temperature spin glass like phase of interacting particle
systems has been revealed from measurements of dc-magnetic relaxation
\cite{Djurberg,Jonsson2a} and relaxation of the low frequency
ac-susceptibility. \cite{Mamiya2}  In this paper, the non-equilibrium
dynamics in the spin glass like phase is further elucidated and
special interest is put into a dynamic memory effect, which has
recently been observed in spin glasses.
\cite{Jonason,Jonsson2,Jonason2,Djurberg2}  
In a ``memory experiment'' the ac-susceptibility is measured at a low frequency, and the cooling of the spin glass sample is halted at one (or more) temperatures. 
During a halt, $\chi$ slowly decays, but when cooling is resumed $\chi$
gradually regains the amplitude of a continuous cooling experiment.
Upon the following heating, $\chi$ shows dips at the temperatures
where the temporary halts occurred; a memory of the cooling history has been imprinted in the spin structure. (For an illustration see Fig.~1 in Ref.\onlinecite{Jonsson2}.)  A similar memory of the cooling
process has recently been reported in an interacting nano-particle
system.  \cite{PJ}

\section{Theoretical background}
For spin glasses, the nonequilibrium dynamics has been interpreted
within a phenomenological real space model, \cite{Jonsson2} adopting
important concepts from the droplet model.\cite{FisherHuse} We will
use the same real space model to discuss similarities and
dissimilarities between the low temperature phase of interacting
particle systems and that of spin glasses.

The droplet model was derived for a short-range Ising spin glass, but
important concepts like domain growth, chaos with temperature, and
overlap length should be applicable also for particle systems
exhibiting strong dipole-dipole interaction and random orientations
of the anisotropy axes.
Chaos with temperature means that a small temperature shift changes
the equilibrium configuration of the magnetic moments completely on
sufficiently long length scales.  The length scale, up to which no
essential change in configuration of the equilibrium state is observed
after a temperature step $\Delta T$, is called the overlap length
$l(\Delta T)$.  The development towards
equilibrium is governed by the growth of equilibrium domains.  The
typical domain size after a time $t_{\mathrm w}$ at a constant
temperature $T$ is
\begin{equation}
R(T,t_{\mathrm w}) \propto \left(\frac{T \ln(t_{\mathrm
w}/\tau)}{\Delta(T)}\right)^{1/\psi},
\label{R}
\end{equation}
where $\tau$ is the relaxation time of an individual magnetic moment,
$\Delta(T)$ sets the free energy scale and $\psi$ is a barrier
exponent.  For spin glasses, the atomic relaxation time is of the
order of $10^{-13}$ s independent of temperature, while for magnetic
particles, the individual particle relaxation time is given by an
Arrhenius law as
\begin{equation}
\tau = \tau_0 \exp \left(\frac{KV}{k_B T}\right),
\label{tau}
\end{equation}
where $\tau_0 \sim 10^{-12} - 10^{-9}$ s, $K$ is an anisotropy constant,
$V$ the particle volume and thus $KV$ the anisotropy energy barrier.  
Due to the inevitable polydispersivity of particle systems
the individual particle relaxation time will also be
distributed, and due to the exponential factor this distribution will 
be broader than the distribution of anisotropy energy barriers.

In the droplet model there exists only two degenerate equilibrium spin
configurations, $\Psi$ and its spin reversal counterpart
$\tilde{\Psi}$.  It is thus possible to map all spins to either of the
two equilibrium configurations.  Let us now consider a spin
glass quenched to a temperature $T_1$ lower than the spin glass 
transition temperature $T_g$.  
At $t=0$ the spins will
randomly belong to either of the two equilibrium states, resulting in
fractal domains of many sizes.  The subsequent equilibration process
at $t>0$ is governed by droplet excitations yielding domain growth to
a typical length scale, which depends on temperature and wait time
$t_{\mathrm w_{1}}$ according to Eq.(\ref{R}).  After this time, fractal
structures typically smaller than $R(T_1,t_{\mathrm w_{1}})$ have become
equilibrated while structures on longer length scales persist.
If the system thereafter is quenched to a lower temperature $T_2$, the
spins can be mapped to the equilibrium configuration at $T_2$ yielding
a new fractal domain structure.  The domain structure achieved at
temperature $T_1$ still fits to the equilibrium configuration at $T_2$
on length scales smaller than the overlap length $l(T_1-T_2)$.  The domain
growth at $T_2$ will start at the overlap length, increase with wait time
$t_{\mathrm w_{2}}$, and end up at the size $R(T_2,t_{\mathrm w_{2}})$.
Heating the system back to $T_1$, the fractal domain growth that
occurred at $T_2$ has only introduced a new and dispersed domain
structure at $T_1$ on length scales $l(T_{1}-T_{2})\leq R \leq
R(T_2,t_{\mathrm w_{2}})$. Note that the large length scale structures, $R
\geq R(T_1,t_{\mathrm w_{1}})$, essentially persist.

The time-dependent response $m(t_{\mathrm obs})$ to a weak magnetic
field applied at $t_{\mathrm obs} = 0$, is due to a continuous
magnetization process, governed by polarization of droplets of size
\begin{equation}
L(T,t_{\mathrm obs}) \propto \left(\frac{T \ln(t_{\mathrm
obs}/\tau)}{\Delta(T)}\right)^{1/\psi}.
\label{L}
\end{equation}
Since $L(T,t_{\mathrm obs})$ grows with the same logarithmic rate as
$R(T,t_{\mathrm w})$, the relevant droplet excitations and the actual
domain sizes become comparably large at time scales $\ln t_{\mathrm obs}
\approx \ln t_{\mathrm w}$.  For $\ln t_{\mathrm obs} \ll \ln t_{\mathrm
w}$ the
relevant excitations occur mainly within equilibrated regions, while
for $\ln t_{\mathrm obs} \gg \ln t_{\mathrm w}$ these excitations
occur on
length scales of the order of the growing domain size, and involves domain walls, yielding a
nonequilibrium response.  A crossover from equilibrium to
nonequilibrium dynamics occurs for $\ln t_{\mathrm obs} \approx \ln
t_{\mathrm
w}$, seen as a maximum (or only a bump) in the relaxation rate $S(t)
= h^{-1}
\partial m(t) / \partial \ln t$ vs. $\ln t$ curves.

Dc-relaxation and ac-susceptibility experiments are related
through the relations: $M(t) \approx \chi'(\omega)$ and $S(t) \propto
\chi''(\omega)$, when $t = 1/\omega$.\cite{Lundgren}  
Hence, the aging behavior is also
observed in low frequency ac-susceptibility measurements.  Different
from dc-relaxation measurements is that the observation time is
constant, $t_{\mathrm obs}=1/\omega$, implying that the probing
length scale $L(T,1/\omega)$ is fixed for a given temperature.

\section{Experimental}
The sample consisted of ferromagnetic particles of the amorphous
alloy Fe$_{1-x}$C$_x$ ($x\approx$0.22) prepared by thermal
decomposition of Fe(CO)$_5$ in an organic liquid (decalin) in the
presence of surfactant molecules (oleic acid) as described in Ref.
\onlinecite{Wonterghem}. After the preparation, the carrier liquid was
evaporated and the particles were transferred to xylene in an
oxygen-free environment resulting in a ferrofluid with a particle
concentration of 5 vol\% (determined by atomic absorption
spectroscopy). The particles were separated by the surfactant
coating and could only interact via magnetic dipole-dipole
interactions. During measurements, the ferrofluid was contained in
a small sapphire cup sealed with epoxy glue to prevent
oxidation of the particles. The sample was only measured and
exposed to magnetic fields well below the melting point of xylene
($\approx$ 248 K). A droplet of the ferrofluid was dripped onto a
grid for transmission electron microscopy and was left to oxidize
prior to the study. The electron micrographs revealed a nearly
spherical particle shape. The volume-weighted size distribution
(after correction for the change in density due to oxidation) was
described well by the log-normal distribution, $f(V)dV
=(\sqrt{2\pi}\sigma_VV)^{-1}\exp[-\ln^2(V/V_{m})/(2\sigma_V^2)]dV$,
with the median volume $V_{m}= 8.6\cdot 10^{-26}$ m$^3$ and
logarithmic standard deviation $\sigma_{V}=0.19$. These parameters
correspond to a log-normal volume-weighted distribution of
particle diameters with the median value $d_{m}=(6V_{m}/\pi)^{1/3}
= 5.5$ nm and $\sigma_d = \sigma_V/3 = 0.062$. This particle size
is slightly larger and the value of $\sigma_V$ is slightly lower
than those reported in recent studies
\cite{Djurberg,Jonsson,Hansen98} performed on a different
batch of particles.
The 5 vol\% sample has been subject to a detailed study of the
correlations and magnetic dynamics using a wide range of
observation times and temperatures. \cite{Hansen99} 
Collective behavior is observed at temperatures below 40~K.

A non-commercial low-field SQUID magnetometer \cite{SQUID} was used
for the measurements.  All ac-susceptibility measurements were
performed at a frequency of 510 mHz and a rms value of the ac-field
of 0.1~Oe.  The cooling and heating rate was 0.25~K/min.  For the
relaxation measurements the dc field was 0.05~Oe, while the background
field was less than 1~mOe.

\section{Results and Discussion}
\subsection{General behavior}
\begin{figure}[htb]
\centerline{\epsfig{figure=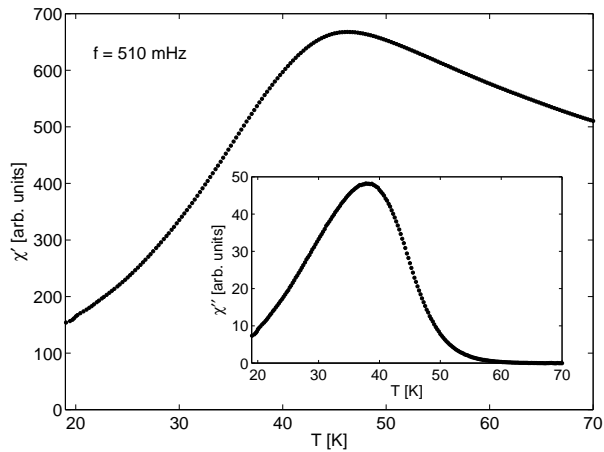,width=8cm}}
\caption[]{The ac-susceptibility vs. temperature at $f=510$ mHz.}
\label{susc}
\end{figure}

\begin{figure}[htb]
\centerline{\epsfig{figure=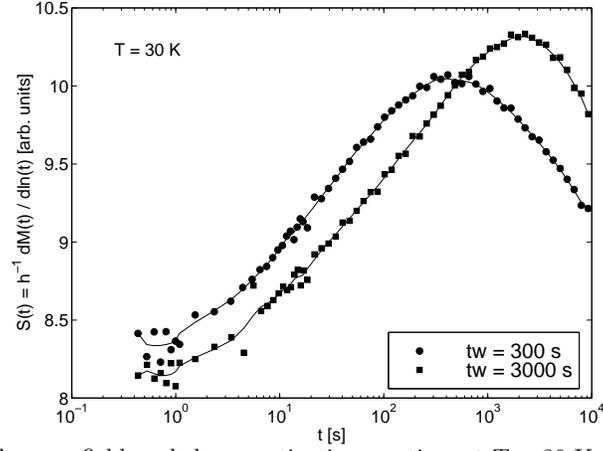,width=8cm}}
\caption[]{The relaxation rate of the zero field cooled magnetization
vs. time at $T=30$~K. The wait time is 300 s (circles) and 3000 s
(squares).}
\label{aging}
\end{figure}

Fig.~\ref{susc} shows the ac susceptibility of the particle sample
measured while cooling.  The onset of the out-of-phase component of
the susceptibility is less sharp than for spin glasses.  
Nevertheless, there exists collective spin glass like dynamics in the particle system, which can be studied by magnetic aging experiments.  
The `conventional' aging experiment is a
measurement of the wait time dependence of the response to a field
change, after either field cooling (thermoremanent magnetization, TRM)
or zero field cooling (ZFC).  In Fig.~\ref{aging}, the relaxation rate
from two ZFC magnetic relaxation measurements is shown. The particle
system was directly cooled from a high temperature to 30~K where it
was kept a wait time of 300 and 3000~s, respectively, before the
magnetic field was applied.  A clear wait time dependence is seen; the
relaxation rate displays a maximum at $\ln t_{\mathrm obs} \approx \ln
t_{\mathrm w}$.  Both particle systems and spin glasses exhibit
magnetic aging that affects the response function in a similar way,
but the wait time dependence appears weaker for the particle system.
Comparing the ratio between the magnitude of $S(t)$ at the maximum and
at the short time limit (0.3 s in our current experiment) we find a
value of order 1.3 in the current sample, whereas most spin glasses
would show a value $\geq$ 2, at a corresponding temperature and wait
time.  The illustrated differences between the dynamics of a particle
system and an archetypical spin glass can be assigned to the wide
distribution of particle relaxation times and that some particles
may relax independently of the collective spin glass phase due to 
sample inhomogeneities.

\begin{figure}[htb]
\centerline{\epsfig{figure=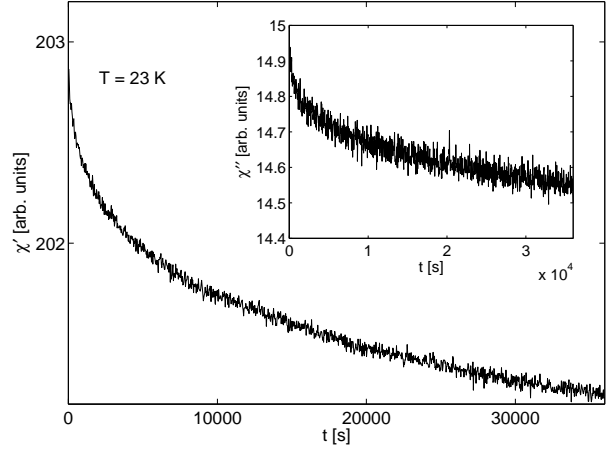,width=8cm}}
\caption[]{Ac-susceptibility vs. time at $T=23$~K. Same units as in
Fig~\ref{susc},  $f=510$ mHz.}
\label{ac-relax}
\end{figure}

The relaxation in the ac-susceptibility, after cooling the sample to
23 K with the same cooling rate as in the reference measurement, is
shown in Fig.~\ref{ac-relax}.  The relaxation in absolute units is
larger for $\chi'$ than for $\chi''$ and, in contrast to spin glasses
\cite{Jonason}  where the aging phenomena are best exposed in
$\chi''$, the effects of aging are more clearly seen in $\chi'$ 
for this particle system.
Measuring the ac-susceptibility with slower cooling rates gives
lower values of the ac-susceptibility.
The effect of a slower cooling rate is largest just below the
transition temperature where the effect
of aging also is largest.

\subsection{Temperature cycling}
\label{Cycl}
One experimental procedure that has been used for spin glasses to
confirm the overlap length concept is temperature cycling.
\cite{Sandlund} The measurements are similar to conventional aging
experiments, but after the wait time $t_{\mathrm w_{1}}$ at the
measuring temperature $T_{\mathrm m}$, a temperature change $\Delta T$
is made, and the system is exposed to a second wait time $t_{\mathrm
w_{2}}$ before changing the temperature back to $T_{\mathrm m}$, where
the field is applied and the magnetization is recorded as a function
of time.

\begin{figure}[htb]
\centerline{\epsfig{figure=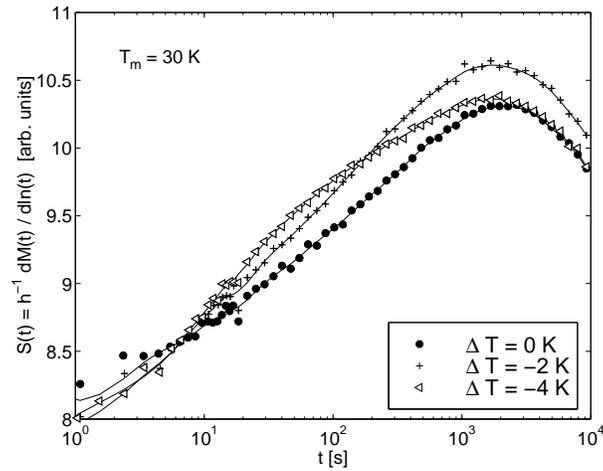,width=8cm}}
\caption[]{Relaxation rate $S(t)$ vs. time. The sample has been
aged $t_{\mathrm w_{1}}=3000$~s at $T_{\mathrm m}=30$~K and afterward
subjected to a negative temperature cycling $\Delta T$ during
$t_{\mathrm w_{2}}= 10 000$~s, immediately prior to the
application of the field $H= 0.05$~Oe; Circles: $\Delta T = 0$;
Pluses $\Delta T
= -2$~K; Triangles:  $\Delta T = -4$~K.}
\label{Ncycl}
\end{figure}

Fig.~\ref{Ncycl} shows results from measurements with negative
temperature cycling, at $T_{\mathrm m}= 30$~K, using $t_{\mathrm w_{1}}$=
3000 and $t_{\mathrm w_{2}}$=10000 s.  For $0 > \Delta T \gtrsim -2$~K
the additional wait time at the cycling temperature mainly shifts the
maximum in the relaxation rate to longer times.
The shift is largest for the smallest $\Delta T$ and the maximum
returns continuously towards $t_{\mathrm w_{1}}$ as $\Delta T$ is increased.
For $\Delta T = -2$~K, the maximum appears
at $\ln t \approx \ln t_{\mathrm w_{1}}$ and the relaxation rate behaves
rather similar to the curve without the cycling only showing a
somewhat larger magnitude (see Fig~\ref{Ncycl}).  For a larger
temperature step, $\Delta T = -4$~K, the relaxation
rate is enhanced at times shorter than $t_{\mathrm w_{1}}$, while the
relaxation appears unaffected at longer time scales.  The behavior can
be interpreted in terms of an interplay between the domain growth and
the overlap length: for small temperature steps, the domain size
$R_1(T_{\mathrm m},t_{\mathrm w_{1}})$ attained during $t_{\mathrm w_{1}}$
at $T_{m}$ is shorter than the overlap length at the cycling
temperature and the domain growth proceeds essentially unaffected by
the temperature change, only with a slower rate.  For sufficiently
large temperature steps the overlap length is shorter than
$R_1(T_{\mathrm m},t_{\mathrm w_{1}})$ and the domain growth at the lower
temperature creates a new domain structure on short length scales,
$R_2(T_{\mathrm m}+\Delta T,t_{\mathrm w_{2}})$.  For the temperatures and
wait times used in the experiments $R_2(T_{\mathrm m}+\Delta
T,t_{\mathrm w_{2}}) < R_1(T_{\mathrm m},t_{\mathrm w_{1}})$.  Returning to
$T_{\mathrm m}$, the new $R_{2}$ structures do not overlap with the
equilibrium configuration and yield the apparent non-equilibrium
nature of the dynamics on short time scales (cf. Fig.~\ref{Ncycl} for $\Delta T = -4$~K).

\begin{figure}[htb]
\centerline{\epsfig{figure=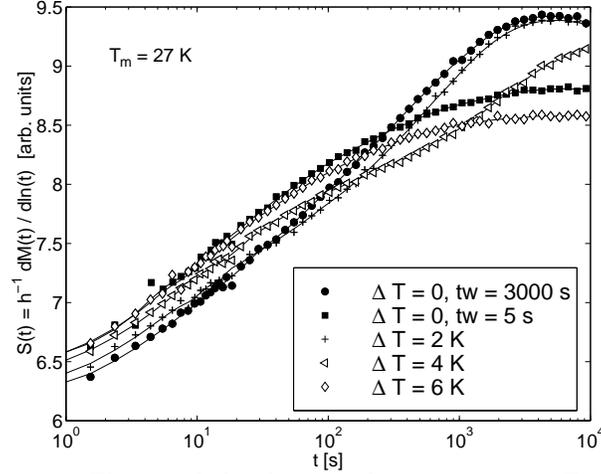,width=8cm}}
\caption[]{Relaxation rate $S(t)$ vs. time. The sample has been
aged $t_{\mathrm w_{1}}=3000$~s at $T_{\mathrm m}=27$~K and afterward
subjected to a positive temperature cycling $\Delta T$ during a time
$t_{\mathrm w_{2}}$, immediately prior to the application of the field
$H=
0.05$~Oe; Circles: $\Delta T = 0$; Pluses $\Delta T = 2$~K,
$t_{\mathrm w_{2}}=5$~s; Triangles:  $\Delta T = 4$~K, $t_{\mathrm
w_{2}}=5$~s; Diamonds: $\Delta T = 6$~K, $t_{\mathrm w_{2}}=30$~s.
Squares are data obtained after the sample has been aged at 27~K for 5~s only.}
\label{Pcycl}
\end{figure}

Measurements with positive temperature cycling with $T_{\mathrm m}=
27$~K are shown in Fig.~\ref{Pcycl}.  For small temperature steps
($\Delta T \lesssim 2$~K) the maximum of the relaxation rate remains
at $\ln t \approx \ln t_{\mathrm w_{1}}$, but for larger temperature
steps it is seen that the relaxation rate approaches the relaxation
rate of an aging measurement with short wait time.  This can again be
explained by the overlap length becoming shorter than $R_1(T_{\mathrm
m},t_{\mathrm w_{1}})$ when $\Delta T \gtrsim 2$~K.  
Since the domain growth rate increases fast with increasing temperature, the
system will then look more and more reinitialized when returning to
$T_{\mathrm m}$ after larger $\Delta T$ and/or longer $t_{\mathrm w_{2}}$.

\subsection{Memory}
\begin{figure}[htb]
\centerline{\epsfig{figure=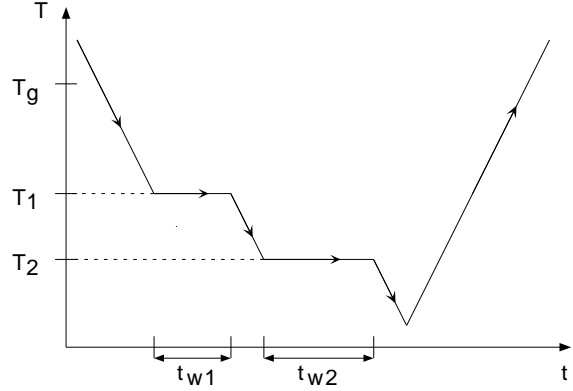,width=8cm}}
\caption[]{The experimental procedure in an ac-susceptibility memory
experiment with two temporary halts during cooling.}
\label{memproc}
\end{figure}

\begin{figure}[htb]
\centerline{\epsfig{figure=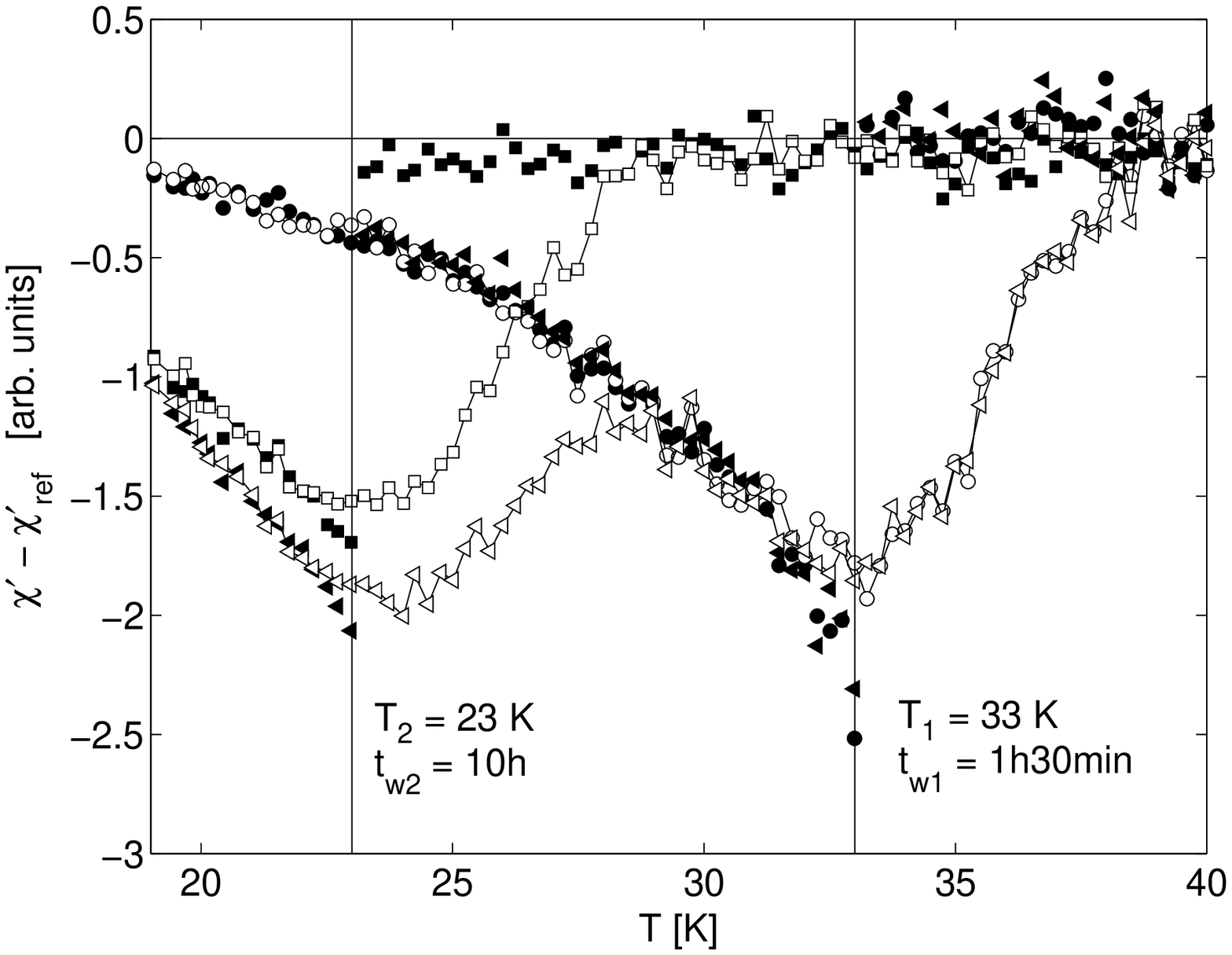,width=8cm}}
\caption[]{$\chi'(T)-\chi'_{\mathrm ref}(T)$ vs. $T$ measured while
cooling (solid symbols) or heating (open symbols). Circles, the
cooling was halted at 33~K for 1~h~30~min; squares, the cooling was
halted at
23~K for 10h and triangles, the cooling was halted at $T_1 = 33$~K
for $t_{\mathrm w_{1}} = 1$~h~30~min and at $T_2 = 23$~K for $t_{\mathrm
w_{1}} = 10$~h. Units and frequency are the same as in
Fig.~\ref{susc}}
\label{memory1}
\end{figure}

The experimental procedure for a memory experiment is illustrated in
Fig.~\ref{memproc}.
$\chi'(T)$ is recorded during cooling, employing different halts at
constant temperature, followed by continuous heating.
Fig.~\ref{memory1} shows results from three different memory
experiments:
the first with a single temporary halt at 33 K for 1 h and 30 minutes,
the second with a single
temporary halt at 23 K for 10 h, and the third with temporary halts at
both 33 K and 23 K. The exposed curves show the difference between the
measured and a reference curve.  Reference curves are taken from a
measurement with continuous cooling followed by continuous heating.
Similar to spin glasses, a memory of the cooling history is observed
on heating for both one and two temporary halts.  It is seen that
on the low temperature side, the susceptibility approaches the
reference level slower than on the high temperature side.  This
asymmetry of the dip is more pronounced in the particle system
than for spin glasses.  The relaxation at 23 K after a temporary
halt at 33 K is equally large as the relaxation at 23 K without
earlier halt, except that it starts at a
lower level.  In fact, $\chi'-\chi'_{\mathrm ref}$, measured on
heating for the experiment with two temporary halts at 33~K and 23~K,
is just the sum of $\chi'-\chi'_{\mathrm ref}$ of the two experiments
with a single temporary halt at 33~K and 23~K, respectively.

\begin{figure}[htb]
\centerline{\epsfig{figure=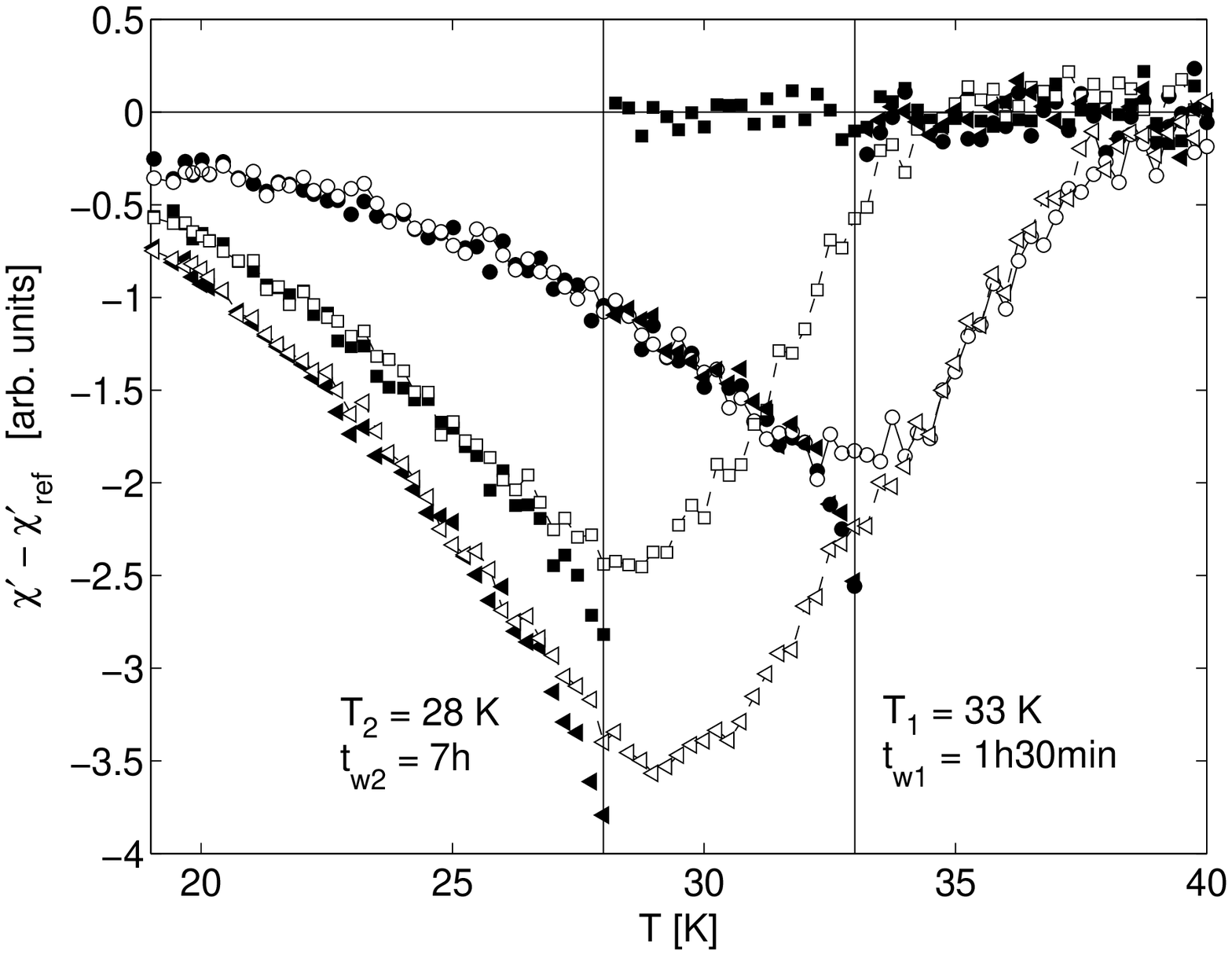,width=8cm}}
\caption[]{$\chi'(T)-\chi'_{\mathrm ref}(T)$ vs. $T$ measured while
cooling (solid symbols) or heating (open symbols). Circles, the
cooling was halted at 33~K for 1h30min; squares, the cooling was
halted at
28~K for 7~h and triangles, the cooling was halted at $T_1 = 33$~K
for $t_{\mathrm w_{1}} = 1$~h~30~min and at $T_2 = 28$~K for $t_{\mathrm
w_{1}} = 7$~h.
Units and frequency are the same as in
Fig.~\ref{susc}}
\label{memory2}
\end{figure}

In Fig.~\ref{memory2} is shown a memory experiment similar to the one
in Fig.~\ref{memory1} except that the two temporary halts are closer in
temperature.  The ac-susceptibility has been measured with temporary
halts on cooling at 33~K for 1h30min and at 28~K for 7~h.  On heating,
the double halt experiment only shows one dip.  However, this dip
($\chi'-\chi'_{\mathrm ref}$) still constitutes the sum of
the two heating curves with one halt, as was the case for the
experiments shown in Fig.~\ref{memory1}.  Also for this experiment,
the ac-relaxation curve at the lower temperature is the same with and
without temporary halt at 33~K, except for a constant.

\begin{figure}[htb]
\centerline{\epsfig{figure=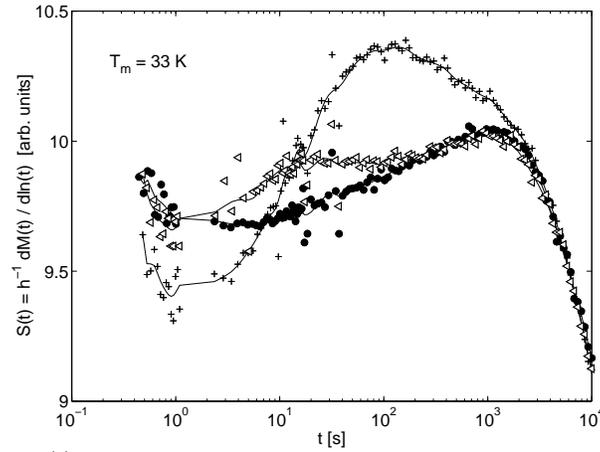,width=8cm}}
\caption[]{Relaxation rate $S(t)$ vs. $\log(t)$. The sample has been
aged for 1h30min at $T_{\mathrm m}=33$~K  and afterward subjected to
a negative temperature cycling $\Delta T$ during
$t_{\mathrm w_{2}}$. Circles: $\Delta T = 0$; Pluses: $T_{\mathrm m}
+\Delta T= 28$~K, $t_{\mathrm w_{2}}=7$~h; Triangles: $T_{\mathrm m}
+\Delta T= 23$~K, $t_{\mathrm w_{2}}=10$~h. $H= 0.05$~Oe.}
\label{dc-relax33}
\end{figure}

We have performed supplementary experiments to the ones shown in
Fig.~\ref{memory1} and \ref{memory2} by measuring the dc-relaxation.
In Fig.~\ref{dc-relax33}, the relaxation rate of three dc-relaxation
curves measured at 33~K are shown.  In the three experiments the
system was quenched to 33~K and then aged for 1h30min.  In one of the
measurements the relaxation was measured immediately after this wait
time, but in the two other measurements a negative temperature cycling
was performed to 23~K for 10~h and to 28~K for 7~h, respectively.  The
dynamics is sizeably affected by the temperature cycling to 28~K, but
only weakly affected by the temperature cycling to 23~K. The observed
increase of the relaxation rate indicates that the equilibrium
structure created at 33~K is erased on short length scales after the
temperature cycling, as discussed in Sec.~\ref{Cycl}.  On time scales
shorter than 10 s, $S(t)$ is even lower for the experiment with a
temperature cycling to 28~K than for the other two experiments.  We
conclude that there is an overlap between the equilibrium
configuration of the magnetic moments at 33~K and 28~K on length
scales shorter than $L(33~{\mathrm K},10~{\mathrm s})$.  Let us now
compare these dc-measurements with the corresponding ac-measurements.
The in-phase component of the ac-susceptibility at a frequency of
510~mHz records the integrated response corresponding to observation
times $1/\omega \approx$ 0.3~s and shorter.  In the ac-experiment with
two temporary halts at 33 and at 23~K (Fig.~\ref{memory1}), no effect of
the aging at 23~K is seen when heating through 33~K. However, $\chi'(33~{\mathrm K})$
measured on heating for the ac-measurement with temporary halts at
33 and 28~K is lower than $\chi'(33~{\mathrm K})$ for the measurement with
only a single halt at 33~K. A lower level of $\chi'(T)$ indicates a
more equilibrated system and thereby also a lower relaxation rate.
This supports
the above conclusion of some overlap between the equilibrium
configurations at 28~K and 33~K on the length scales seen by
observation times shorter than 0.3~s.

\begin{figure}[htb]
\centerline{\epsfig{figure=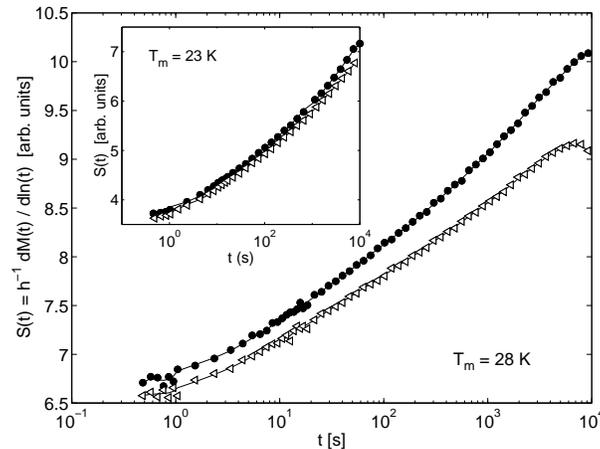,width=8cm}}
\caption[]{Relaxation rate $S(t)$ vs. $\log(t)$. The sample has been
aged, after a quench, for $t_{\mathrm w}=7$~h at $T_{\mathrm m}=28$~K
(circles) and after a quench with a temporary halt at 33~K for
1h30min (triangles), $H= 0.05$~Oe. Inset: $T_{\mathrm m}=23$~K and
$t_{\mathrm w}=10$~h.}
\label{dc-relax28}
\end{figure}

Fig.~\ref{dc-relax28} shows the relaxation rate measured at 28 and
23~K after a quench to $T_{\mathrm m}$ followed by a wait time of
7~h and 10~h, respectively.  The same measurements except for a
temporary halt at 33~K for 1h30min are also shown.  The relaxation
rate is lower for the measurement with a  temporary halt at 33~K
for both temperatures, even though the difference between a temporary
halt and a direct quench is larger at 28~K. These measurements,
as the corresponding ac-measurements, show that there is some
overlap on short length scales between the equilibrium configuration
at 33~K and all lower temperatures for which we have measured the
ac-susceptibility.

\section{Conclusions}
The non-equilibrium dynamics in the low temperature spin glass like
phase of an interacting Fe-C nano-particle sample is found to largely
mimic the corresponding spin glass dynamics.  The observed differences
may be accounted for by the strongly temperature dependent and widely
distributed relaxation times of the particle magnetic moments compared
to the temperature independent and monodispersed relaxation times of
the spins in a spin glass.  Within a droplet scaling picture of a
particle system, these factors strongly affects the associated length
scales and growth rates of the domains and droplet excitations as well
as the roughness of the domain walls.  In addition, there might exist
individual particles that do not take part in the collective low
temperature spin glass phase but relax independently.

\acknowledgements
This work was financially supported by The Swedish Natural Science Research
Council (NFR).

\end{document}